\documentclass[amsmath,amssymb,prl,aps,showpacs,twocolumn,10pt]{revtex4-1}
\usepackage{amsmath,amsfonts,amssymb,amsthm,graphics,graphicx,epsfig,bbm}
\usepackage[colorlinks=true,citecolor=blue,linkcolor=red,urlcolor=blue]{hyperref}
\usepackage[usenames]{color}
\usepackage{graphicx}
\usepackage{subfigure}
\usepackage{amsmath}
\usepackage{epsfig}
\usepackage{dcolumn}
\usepackage{bm}
\usepackage{color}
\usepackage{epstopdf}
\usepackage{amssymb}
\usepackage{amstext}
\usepackage{latexsym}
\usepackage{hyperref}
\usepackage{amsfonts}
\usepackage{psfrag}
\usepackage{xcolor}
\usepackage[normalem]{ulem}
\usepackage{dsfont}
\usepackage{txfonts}
\usepackage{overpic}
\usepackage{ifthen}

\newcommand{\an}[2]{\ifthenelse{\equal{#1}{}}{\ensuremath{\hat{#1}_{#2}}}{\ensuremath{\hat{#1}^{\protect\phantom{\dagger}}_{#2}}}}

\begin{document}

\title{Topological Thouless pumping in arrays of coupled spin chains}

\author{V. M. Bastidas}
\email{victor.bastidas.v.yr@hco.ntt.co.jp}

\affiliation{%
NTT Basic Research Laboratories \& Research Center for Theoretical Quantum Physics,  3-1 Morinosato-Wakamiya, Atsugi, Kanagawa, 243-0198, Japan}

\date{\today}

\begin{abstract}
Thouless pumping is a mechanism to perform topologically protected transport of particles by adiabatically modulating the Hamiltonian. The transported current is a topological invariant that is intimately related to the integer quantum Hall effect. Most of the previous works focus on topological pumping in linear and square lattices. 
In this work, we theoretically propose a mechanism to perform topological pumping in arrays of  spin chains with complex geometries. 
To achieve this, we consider an array where the spin chains are coupled through their edges, which allows to split the populations to generate superpositions of spin excitations in different spin chains. 
We show that due to the topological protection, the quantum superpositions can be transported through the array against the effect of  disorder.
This approach will open a new avenue to transport excitations and correlated states with potential applications in quantum technologies and information processing. Our ideas can be realized in state-of-the-art quantum simulators such as cold atoms and superconducting qubit arrays.
\end{abstract}

\maketitle

{\it{Introduction.---} }Transport phenomena play a very important role in diverse fields and in technological applications~\cite{Beenakker1997,kohler2005driven,Hanggi2009,Estarellas2017,Engelhardt2018,Engelhardt2019}. Even in fully controllable systems such as quantum simulators~\cite{Georgescu2014}, the transport can be affected by local imperfections~\cite{Ronke2016,Gong2021}. The observation of integer quantum Hall effect was the first demonstration of robust transport protected by topology~\cite{Klitzing1980,vonKlitzing1986,von202040}. Motivated by the chiral edge currents characteristic of quantum Hall effect, Thouless proposed a mechanism to transport particles along a one-dimensional lattice by adiabatically modulating the Hamiltonian of a lattice~\cite{thouless1983,Thouless1984,Seiler1985,Kwek2014,Gong2015}. This is currently known as topological Thouless pumping and allows unidirectional transport of particles with topological protection against imperfections. 

Thouless pumping was originally proposed in condensed matter, but currently it is a topic of active theoretical research in diverse communities. There are theoretical proposals for topological pumping in quantum simulators such as in cold atoms~\cite{Taddia2017,haug2019}, arrays of superconducting qubits~\cite{Tangpanitanon2016topological,Mei2018}, and in photonic systems~\cite{Brosco2021}. The idea of Thouless has been extended to allow pumping of energy instead of particles, which is referred to as  Floquet-Thouless energy pump~\cite{Kolodrubetz2018,Friedman2018,Nathan2021}. 
Recent experiments have demonstrated topological pumping in diverse quantum simulators such as  arrays of waveguides~\cite{kraus2012,Silberberg2015,2018Zilberberg} and cold atoms~\cite{bloch2016, takahashi2016,2018Lohse}.  In a recent experiment, ideas from Topological pumping have been used to create synthetic dimensions~\cite{Ozawa2019} in one- and two-dimensional lattices that allow to transport sound in acoustic crystals~\cite{Chen2021}. For technological applications it would be desirable to perform robust transport in arrays with arbitrary geometries. However, in despite of the enormous theoretical and experimental progress, topological pumping in geometries beyond linear and square lattices remains largely unexplored.

\begin{figure*}
	\includegraphics[width=0.9\textwidth]{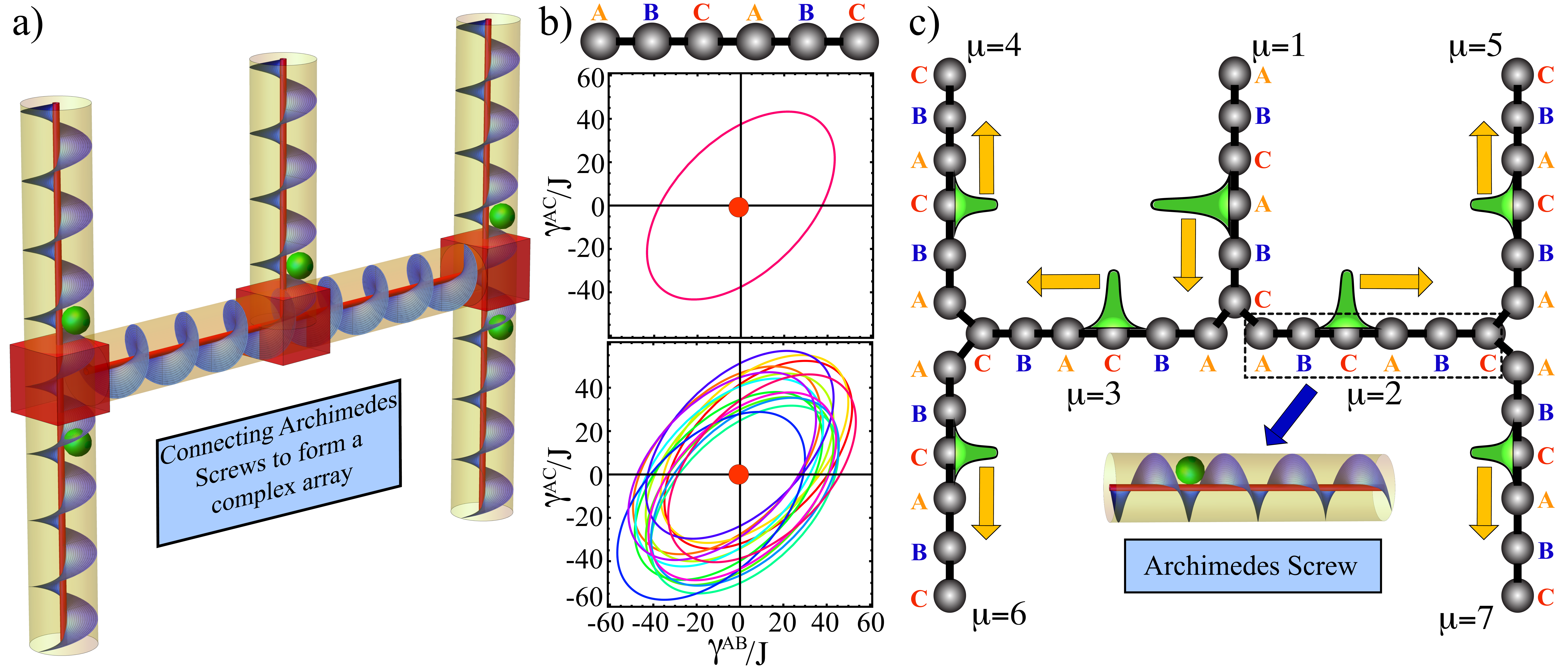}
	\caption{Transporting particles in an array of Archimedes screws. a) Illustrates an array where the links are Archimedes screws and the red boxes are the junctions that allow to connect them. In this configuration, it is possible to carefully tune the rotation of each screw to transport the particles in different directions. b) Shows a one dimensional spin chain with $L=6$ sites that can be seen as a system of coupled trimers. Within a trimer, the sites A, B and C have energies angular frequencies $\omega^{\text{A}}(t)$, $\omega^{\text{B}}(t)$ and $\omega^{\text{C}}(t)$, respectively. The adiabatic modulation of the onsite energies defines a parametric curve $[\gamma^{\text{AB}}(t),\gamma^{\text{AC}}(t)]$, where $\gamma^{\text{AB}}(t)=\omega^{\text{A}}(t)-\omega^{\text{B}}(t)$ and $\gamma^{\text{AC}}(t)=\omega^{\text{A}}(t)-\omega^{\text{C}}(t)$. The middle and lower panels of b) show the parametric curve for the clean system and for many realizations of disorder, respectively. Thouless pumping is possible as long as the curves enclose the origin (red dot). c) Illustrates how to couple trimerized spin chains to build a quantum mechanical analog of the array depicted in a). A localized excitation initially prepared at the spin chain $\mu=1$ can be distributed through the array.}
	\label{Fig1}
\end{figure*}

 In this letter, we propose a way to perform topological pumping in arrays of coupled spin chains with complex geometries. The idea is based on the Archimedes screw from the ancient Greece, which is the mechanical analog of Thouless pumping. Intuitively, it is possible to interconnect several Archimedes screws to form an array as in Fig.~\ref{Fig1}~a). To build a quantum analog of this device, we consider an array of coupled spin chains as shown in Figs.~\ref{Fig1}~b)~and~\ref{Fig1}~c). The interchain couplings allow create superpositions of spin excitations in the array. 
The adiabatic modulation of the onsite energies allows to transport superpositions of spin excitations  [see Fig.~\ref{Fig1}~c)] through the array against imperfections such as disorder. The proposed scheme is intimately related to arrays of quantum Hall systems and can be extended to in order to transport correlated particles in quantum simulators.

{\it{Thouless pumping in a spin chain.---} } Before entering into the details on how to transport particles in an array of spin chains [see Fig.~\ref{Fig1}~(c)], we briefly discuss the general idea of Thouless pumping. To do this, here we focus on a single spin chain as in Fig.~\ref{Fig1}~(b) described by the Hamiltonian
\begin{equation}\label{eq:SpinAAmodel}
\hat{H}^{(\mu)}(t)=\frac{\hbar}{2}\sum^{L}_{l=1}\omega_{\mu, l}(t)\sigma^z_{\mu, l}+ \frac{\hbar J}{2}\sum^{L-1}_{l=1}  ( \sigma^x_{\mu, l}
\sigma^x_{\mu, l+1}+\sigma^y_{\mu, l}
\sigma^y_{\mu, l+1})
\ . 
\end{equation}
The index $\mu$ is introduced to label the spin chain. Here $L$  is the number of spins and $\sigma^a_{\mu, l}$ with $a=x,y,z$ denote the Pauli-matrices at the $l$-th site of the chain. In this work, the onsite angular frequencies  $\omega_{\mu, l}(t) = \Delta \cos[2\pi (l-1)b+\Omega t+\theta_{\mu}]$ are adiabatically modulated with an amplitude $\Delta$. The angular frequency of the modulation is 
 $\Omega$, while $\theta_{\mu}$ and ${T=2\pi/\Omega}$ denote its phase and period, respectively. The parameter $b$ is a real number that is intimately related to the external magnetic field in the integer quantum Hall effect~\cite{1976Hofstadter} and determines the spatial period of the adiabatic modulation. In this work $J$ is the interaction strength between neighboring spins.  

The Hamiltonian~\eqref{eq:SpinAAmodel} is a spin version of the Aubry-Andre model~\cite{aubry80,kraus2012}, which is intimately related to the Hofstadter model of two dimensional electrons in a magnetic field~\cite{1976Hofstadter,aidelsburger2015measuring}. It important to define the notation used in this work. From now on we will restrict ourselves to the single-excitation subspace. The state $|1_{\mu,l}\rangle=|\downarrow\downarrow\cdots\uparrow_{\mu,l}\cdots\downarrow\downarrow\rangle$ denotes a spin flip localized at the $l$-th site of the spin chain. 

Next we briefly discuss a transport mechanism known as topological Thouless pumping~\cite{thouless1983,Thouless1984}. This allows to generate a chiral current of particles by adiabatically deforming the onsite angular frequencies $\omega_{\mu,l}(t)$ in Eq.~\eqref{eq:SpinAAmodel} as a function of time. During a period of the adiabatic modulation the current is proportional to the Chern number~\cite{thouless1983,Thouless1984}, which is follows from the quantization of the conductance in the integer quantum Hall effect~\cite{vonKlitzing1986,von202040}. Here, we consider the case $b=1/3$ so that the spin chain corresponds to a set of $L/3$ coupled trimers labelled by $r=1,\ldots,L/3$. The sites $l=3r-2,3r-1,3r$ within a given trimer are labelled by $A,B$ and $C$ as in Figs.~\ref{Fig1}~b)~and~c). 
This lattice structure gives rise to three separated bands and each one of them is characterized by a Chern number that can be either ${\boldsymbol{C}=2}$ or ${\boldsymbol{C}=-1}$. More specifically, the Chern number measure the number of trimers that the wave packet pass during a period of the pump~\cite{Tangpanitanon2016topological,haug2019}.

 The topological protection against imperfections can be intuitively understood by considering the angular frequencies $\omega_{\mu,r}^{\text{A}}(t)=\Delta \cos[\Omega t+\theta_{\mu}]+\delta_{\mu,r}^{\text{A}}$, $\omega_{\mu,r}^{\text{B}}(t)=\Delta \cos[2\pi/3+\Omega t+\theta_{\mu}]+\delta_{\mu,r}^{\text{B}}$, and $\omega_{\mu,r}^{\text{C}}(t)=\Delta \cos[4\pi/3 +\Omega t+\theta_{\mu}]+\delta_{\mu,r}^{\text{C}}$ within the $r$-th trimer. Here,  $\delta_{\mu,r}^{\text{A}}$,  $\delta_{\mu,r}^{\text{B}},\delta_{\mu,r}^{\text{C}}\in[-W,W]$ denotes local disorder with strength $W$
 for the sites $A,B$ and $C$, respectively. In this work, the disorder is drawn from a uniform distribution. Next consider the parametric curve
 $\Gamma_{\mu,r}(t)=[\gamma_{\mu,r}^{\text{AB}}(t),\gamma_{\mu,r}^{\text{AC}}(t)]$ with $\gamma_{\mu,r}^{\text{AB}}(t)=\omega_{\mu,r}^{\text{A}}(t)-\omega_{r_{\mu}}^{\text{B}}(t)$ and $\gamma_{\mu,r}^{\text{AC}}(t)=\omega_{\mu,r}^{\text{A}}(t)-\omega_{\mu,r}^{\text{C}}(t)$. The curves $\Gamma_{\mu,r}(t)$ are depicted in the middle and lower panels of Fig.~\ref{Fig1} ~b) for the clean system and for several realizations of disorder, respectively. As long as the different curves enclose the origin during the pumping protocol, the transport is topologically protected and robust against imperfections such as disorder. The chiral current generated during this process resembles the Archimedes screw, where geometrical properties of a mechanical device are exploited to perform unidirectional transport. Here, however, the transport is related to topological invariants and hence it is robust against imperfections.  The next step is to find a way to couple the spin chains to build a arrays with arbitrary geometries.
 
 {\it{Building arrays of coupled spin chains.---} } We next show how to interconnect $M$ different spin chains to build arrays with complex geometries. Consider the general Hamiltonian 
\begin{equation}\label{eq:ArchimedesNetwork}
\hat{H}_{\text{Net}}(t)=\sum^M_{\mu=1}\hat{H}^{(\mu)}(t)+ \frac{\hbar}{2}\sum^{M}_{\mu,\nu=1} K_{\mu,\nu} ( \sigma^x_{\mu,\text{ed}}
\sigma^x_{\nu,\text{ed}}+\sigma^y_{\mu,\text{ed}}
\sigma^y_{\nu,\text{ed}})
\ ,
\end{equation}
where $\hat{H}^{(\mu)}(t)$ was previously defined in Eq.~\eqref{eq:SpinAAmodel}. $K_{\mu,\nu}$ denotes the coupling strength between the edges of the $\mu$-th and $\nu$-th spin chains, respectively. Spin operators localized at the edges are denoted by $\sigma^a_{\mu, \text{ed}}$ with $a=x,y,z$. Due to the trimer structure depicted in Fig.~\ref{Fig1}~b), the edges of a given spin chain are always labeled by $A$ and $C$.  In principle, one can connect any number of spin chains to build complex structures as long as the edges $A$ and $C$ of the $\mu$-th and $\nu$-th spin chains are coupled. 
 By coupling the edges of different spin chains, the adiabatic modulation of the onsite energies allows to perform transport in the array. An intuitive way to understand the underlying mechanism is to think of having an array of Archimedes screws as depicted in Fig.~\ref{Fig1}~a). There, the junctions (red boxes) are designed to interconnect the different screws. In this configuration is possible to carefully tune the rotation of the screws in the array to transport the particles in different directions. In the next section, we will provide a more detailed explanation of the transport mechanism in the array of spin chains based on the physics of Thouless Pumping and discuss the role of topological protection.

 \begin{figure}
	\includegraphics[width=0.49\textwidth]{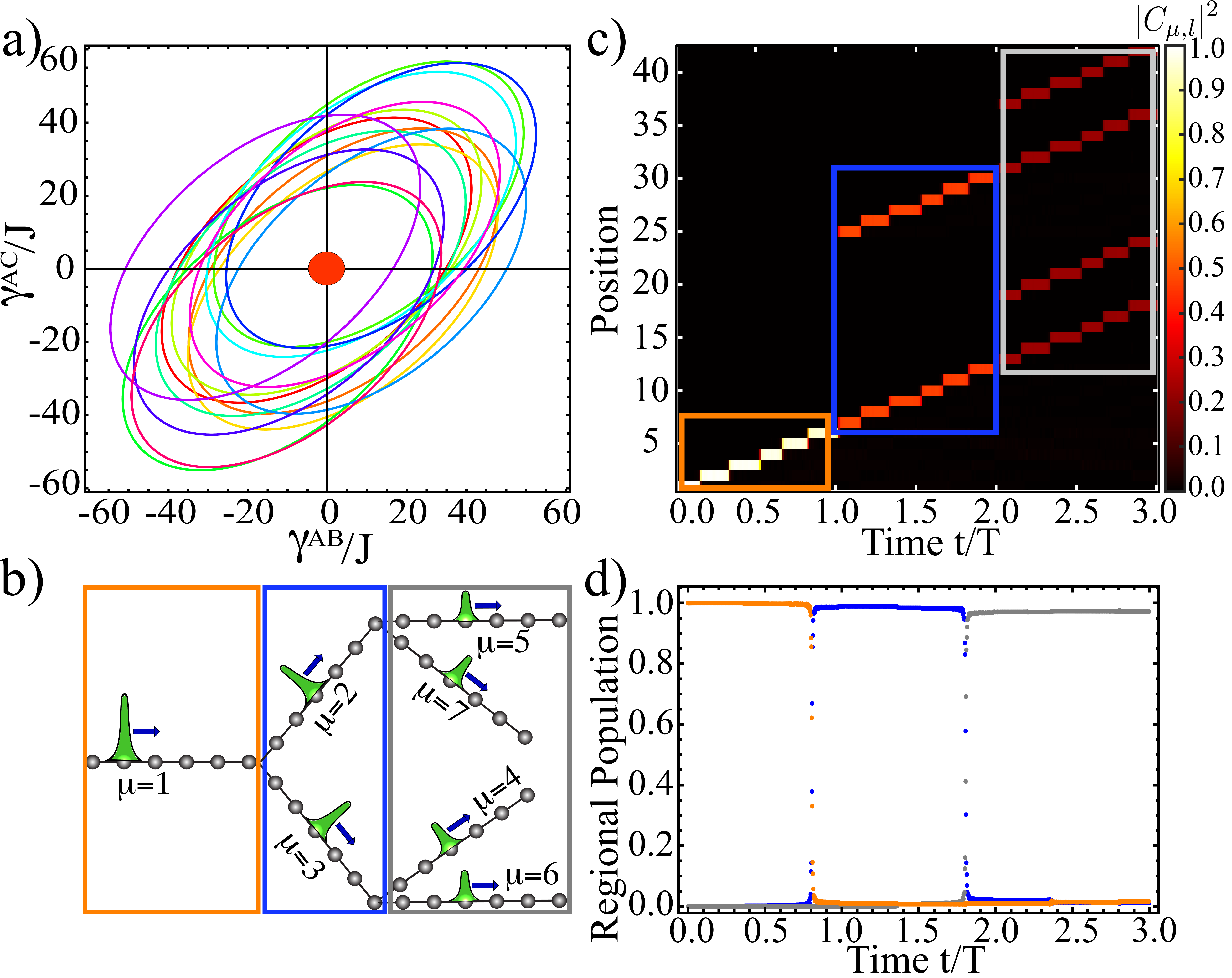}
	\caption{Topological Thouless pumping in an array of $M=7$ coupled spin chains. a) Depicts the curves $\Gamma_{\mu,r}(t)$ for each one of the $ML/3=14$ trimers of the array for a single realization of disorder with strength $W/J=20$. Each curve encloses the origin(red dot). b) Transport of spin excitations in an array with the same geometry as in Fig.~\ref{Fig1}~c). c) Density plot illustrating the quantized transport in  during three periods of the pumping protocol. An excitation is initially prepared at the site A of the spin chain $\mu=1$ and is distributed in the array as depicted in Fig.~\ref{Fig1}~c). The boxes correspond to the regions depicted in b). d) Population dynamics of the regions depicted in b) and c). Here we have chosen $\Delta/J=45$, $\Omega/J=0.015$ and $\theta_{\mu}=\pi/3$. }
	\label{Fig2}
\end{figure}
  
 {\it{Thouless pumping in arrays of spin chains.---} }
 As we discussed in the previous sections, for a given spin chain $\mu$ we consider an adiabatic modulation of the onsite energies $\omega_{\mu,l}(t)$ with a phase $\theta_{\mu}$. This phase can be tuned to transport particles in different directions and velocities, i.e., with a particle current associated to different Chern numbers such as ${\boldsymbol{C}=2}$ or ${\boldsymbol{C}=-1}$. Intuitively, this is similar to changing the rotation direction of each screw in Fig.~\ref{Fig1}~a).  The transport is protected as long as the individual parametric curves $\Gamma_{\mu,r}(t)$ enclose the origin for all the spin chains labelled by $\mu$ as shown in Fig.~\ref{Fig2}~a). In this case, the system remains gapped during the adiabatic modulation of the onsite energies. The gap is responsible for the robustness of the transport against perturbations such as disorder~\cite{Tangpanitanon2016topological,Mei2018}. If the curve does not enclose the origin for a given trimer $r_{\mu}$, this loses the topological protection and becomes a defect that scatters the particles.
 
 So far the discussion is general and can be applied to any geometry of the array. However,  in this work, we provide a concrete example by choosing couplings $K_{1,2}=K_{1,3}=K_{3,5}=K_{3,7}=K_{2,6}=K_{2,4}=J$ between the edges of $M=7$ spin chains as shown in Fig.~\ref{Fig1}~c). Further, we consider a uniform phase $\theta_{\mu}=\pi/3$ for all the spin chains. Note that the edges of the spin chains are always connected in a pattern. For example, the edge $C$ of the chain $\mu=1$ couples to the edges labeled by $A$ from the spin chains $\mu=2$ and $\mu=3$.

 This configuration is designed to transport particles by exploiting the topology of a band with Chern number ${\boldsymbol{C}=2}$, which is the number of trimers that the particle travels to during a period $T$ of the pump~\cite{Tangpanitanon2016topological}.
 Consider an excitation  $|\Psi(0)\rangle=|1_{1,\text{ed}}\rangle$ initially prepared in the edge $A$ of the chain $\mu=1$. After completing the first cycle of the pump, the particle has to be displaced two trimers. During this process it has to cross the edge $C$ [see Fig.~\ref{Fig1}~c)]. This splits the population and allows to create a quantum superposition $|\Psi(T)\rangle\approx \alpha_{\text{T}}(|1_{2,\text{ed}}\rangle+|1_{3,\text{ed}}\rangle)+\beta_{\text{T}} |1_{1,\text{ed}}\rangle$. $|\alpha_{\text{T}}|^2/2$ are the population transmitted to the $A$ edges of the spin chains $\mu=2$ and $\mu=3$. Correspondingly, $|\beta_{\text{T}}|^2$ is a small amount of reflected population due to the interchain coupling. After two periods of the pump, the same mechanism allows to create the superposition state $|\Psi(2T)\rangle\approx\alpha_{\text{2T}}\sum^7_{\mu=4}|1_{\mu,\text{ed}}\rangle+\beta_{\text{2T}}(|1_{2,\text{ed}}\rangle+|1_{3,\text{ed}}\rangle)$. In this case, the transmitted populations to each one of the chains $\mu=4,5,6,7$ are $|\alpha_{\text{2T}}|^2/4$. Some small part of the population is reflected to the chains $\mu=2$ and $\mu=3$. Finally, one can further pump the superposition state until it reaches the edges labelled by $C$ after three periods of the pump. The whole pumping process is illustrated in Fig.~\ref{Fig2}~b). For simplicity, we have refrained from drawing the reflected excitations due to the interchain coupling in Figs.~\ref{Fig1}~c)~and~\ref{Fig2}~b).
An important aspect of our approach is that for any realization of disorder, the interchain coupling via the edge $C$ allows to equally split the transmitted populations. This occurs because the driving phases $\theta_{\mu}$ are uniform in the array. 

There are some related works in the literature that are worth to mention. A recent paper has proposed the idea of a topological beam splitter by creating a topological interface between two Rice-Mele models~\cite{QI2021}. In optics, there is a recent experimental realization of a topological beam splitter in two-dimensional photonic crystals~\cite{He2020}. In this work, we go beyond the idea of topological beam splitters by generalizing the concept of topological pumping to an array of coupled spin chains with arbitrary geometry. In this way, for any realization of disorder, a spin excitation can be equally split to form a quantum superposition in multiple chains. By manipulating the phases $\theta_{\mu}$, one can further change the weights of the quantum superposition. This allows to transport superpositions of spin excitations in a robust fashion against perturbations.

After qualitatively describing the pumping protocol in the array of spin chains, next we will quantitatively describe the dynamics of the populations.
We numerically solve the time-dependent Schr\"{o}dinger equation for the Hamiltonian \eqref{eq:ArchimedesNetwork} of the array and propagate the initial state described above during three periods of the pump.
To keep the dynamics in the adiabatic regime, we consider a pumping frequency $\Omega/J=0.015$. Each spin chain has $L=6$ sites as depicted in Fig.~\ref{Fig2}~b). To demonstrate the robustness of the protocol, we include the effect of strong disorder with strength $W/J=20$. The population dynamics is depicted in Fig.~\ref{Fig2}~c) for a single realization of disorder,  where one clearly sees how the populations split after each period of the pump. 

As a next step, we characterize the back-scattering when the particles reach the edges $C$ of each spin chain. Fig.~\ref{Fig2}~d) shows the dynamics of the regional population inside each one of the three regions of the array depicted in Figs.~\ref{Fig2}~b)~and~~\ref{Fig2}~c).  Our results also show that the back-scattering is mainly due to the interchain coupling. For a single spin chain, the particle moving with Chern number ${\boldsymbol{C}=2}$ is reflected by the edge and moves afterwards with a Chern number ${\boldsymbol{C}=-1}$. When multiple spin chains are coupled, the population can be transferred to other regions of the array, but part of it will be reflected.

\begin{figure}
	\includegraphics[width=0.48\textwidth]{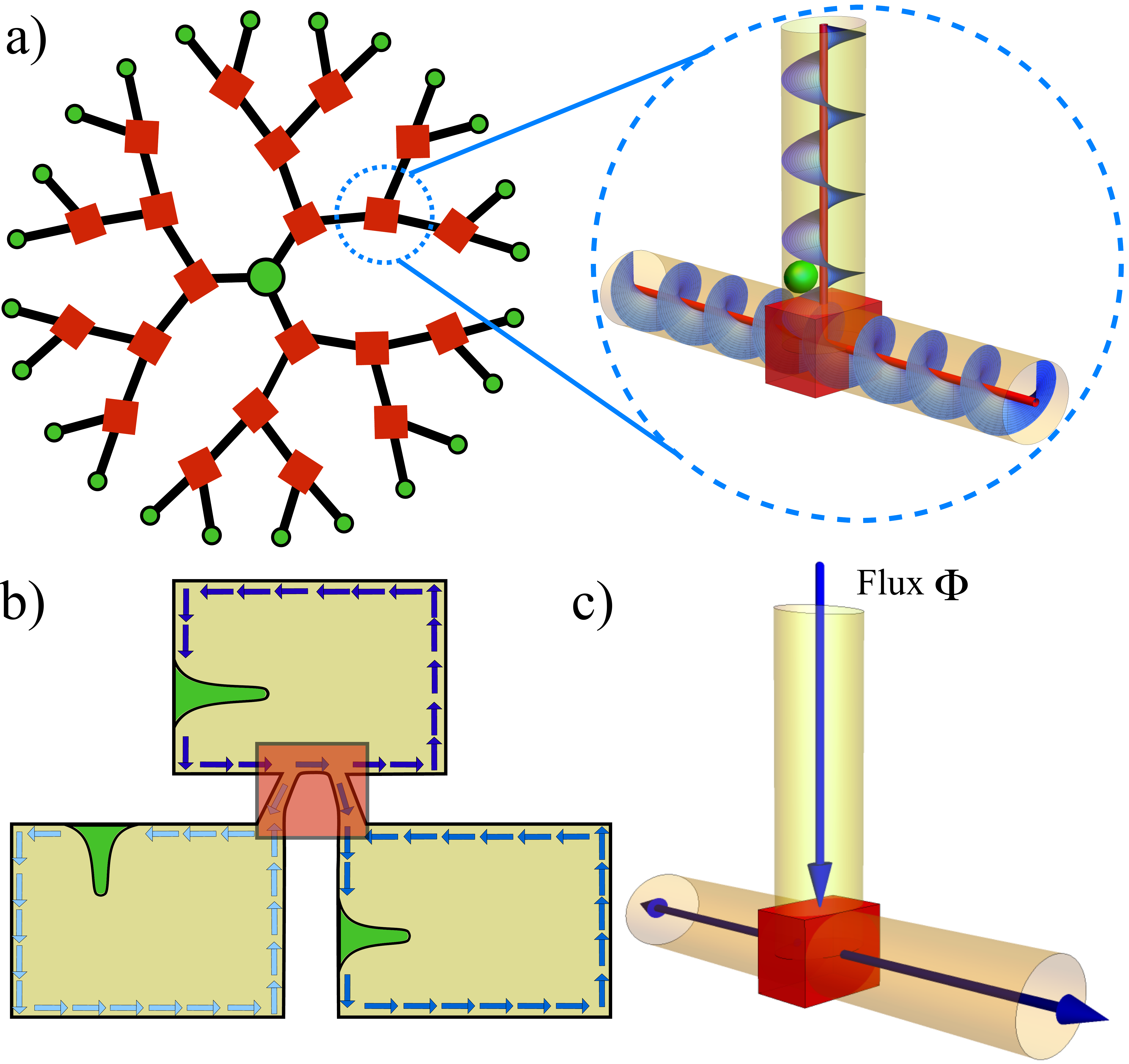}
	\caption{Topological pumping arrays of coupled spin chains and its heuristic relation to the integer quantum Hall effect. a) Depicts an array with complex geometry: A Bethe lattice with coordination number $z=3$. Each link in the depicted network is a spin chain. An initial excitation at the center (green dot) can be pumped to other spin chains of the array against imperfections. The right panel depicts a motif of the network, which can be thought as a junction of Archimedes screws as in Fig.1~\ref{Fig1}~a). b) A Quantum Hall system analog to the motif depicted in a). Here when an edge excitation reaches the (red box), it splits into a superposition of two excitations. Each one of them travels to the edge of a different quantum Hall bar. c) Heuristic interpretation of pumping in the array depicted in b) by using the Laughlin argument in a geometry of interconnected tubes. The blue arrows represent the external magnetic flux.}
	\label{Fig3}
\end{figure}

So far, we have provided numerical results for a concrete example, but our approach can be applied to more complex arrays as the one depicted in Fig.~\ref{Fig3}~a). It is useful to imagine this array as a set of coupled minimal units known as motifs. In this case, the motif allows to split the populations as we discussed above. For pedagogical reasons, it is useful to think of this array as a system of interconnected Archimedes screws. The pumping protocol, allows to transport the initial excitation in the center to the edges of the array [Fig.~\ref{Fig3}~a)].

 {\it{Relation to the integer quantum Hall effect.---} }
In this section we discuss the relation of our results to arrays of Integer Quantum Hall bars. It is well known that topological Thouless pumping was inspired by the integer quantum Hall effect~\cite{thouless1983,Thouless1984}. The explicit relation was pointed out by Laughlin~\cite{Laughlin1981}, who imagined a quantum Hall bar with periodic boundary conditions thread by a time-dependent magnetic flux. The time dependent flux induces a current and due to the Gauge invariance, it should be quantized giving rise to the Chern number~\cite{Laughlin1981}. In turn, one can adiabatically modulate a one dimensional lattice to simulate the dynamics of a two-dimensional Integer quantum Hall bar. In this situation, the adiabatic periodic drive acts as an artificial dimension~\cite{Ozawa2019}.

 Next let us imagine a different type of arrays where two-dimensional quantum Hall systems are coupled as in Fig.~\ref{Fig3}~b). In this case, the chiral edge channel prepared in a Hall bar can tunnel to other bars. The whole setup allows to split the population due to tunneling between edge states. Note that after the excitation crosses the junction, it travels in the opposite direction to the incoming excitation due to chirality. This perspective also fits with our observation of the backscattering discussed above. The system depicted in Fig.~\ref{Fig3}~b) can be considered as a motif. By coupling many of them one can build non-trivial structures to transport edge states. Similarly to the Laughlin argument for the integer quantum Hall effect, the transport of edge states can be interpreted in terms of a current induced by time-dependent flux threading the interconnected tubes depicted in Fig.~\ref{Fig3}~c). The induced current splits at the connection between the tubes.

{\it{Conclusions.---} }
In summary, we consider an array of coupled spin chains under the effect of an adiabatic drive. This allows us to both create quantum superpositions and also to transport them through the array with topological protection against disorder. We show that the interchain couplings allow to split the populations by exploiting the topology of the system. We briefly discussed the relation between our results and arrays of integer quantum Hall systems. Our versatile framework to perform transport in complex geometries can be naturally extended to explore a plethora of fascinating phenomena. For example, in arrays of coupled Ising chains, one can not only transport spin flips, but also kinks and cluster states~\cite{HaugTop2020}. Other future
direction of research is to investigate Hong-Ou-Mandel interference assisted by the Thouless pumping in arrays of coupled chains for a robust transport of entangled states~\cite{Hu2020}. 
It would be interesting to explore arrays of coupled systems with dimension $D>1$. For example, coupled arrays of two-dimensional integer and fractional quantum Hall systems as depicted in Fig.~\ref{Fig3}~b). In the fractional quantum Hall case, our proposed beam splitter may allow to explore Andreev reflections of quasiparticles~\cite{Hashisaka2021}. Our results can be experimentally realized by using quantum simulators such as two-dimensional superconducting quantum processor~\cite{Gong2021}. In this implementation, the frequency of each qubit can be adiabatically modulated to transport spin excitations along different propagation paths with complex geometries.

{\it{Acknowledgments.---} }
I thank H. Bohuslavskyi, M. P. Estarellas, M. Hashisaka and W. J. Munro for fruitful discussions.

\end{document}